\documentclass[pre,superscriptaddress,twocolumn,showpacs]{revtex4}
\usepackage{epsfig}
\usepackage{latexsym}
\usepackage{amsmath}
\begin {document}
\title {Heat conduction and diffusion of hard disks in a narrow channel}
\author{Adam Lipowski}
\affiliation{Faculty of Physics, Adam Mickiewicz University, 61-614
Pozna\'{n}, Poland}
\author{Dorota Lipowska}
\affiliation{Institute of Linguistics, Adam Mickiewicz University,
60-371 Pozna\'{n}, Poland}
\pacs{44.10.+i} \keywords{heat conduction, diffusion, hard disks,
momentum conservation, molecular dynamics}
\begin {abstract}
Using molecular dynamics we study heat conduction and diffusion of
hard disks in one dimensional narrow channels. When collisions
preserve momentum the heat conduction $\kappa$ diverges with the
number of disks $N$ as $\kappa\sim N^\alpha$ $(\alpha \approx
1/3)$. Such a behaviour is seen both when the ordering of disks is
fixed ('pen-case' model), and when they can exchange their
positions. Momentum conservation results also in sound-wave
effects that enhance diffusive behaviour and on an intermediate
time scale (that diverges in the thermodynamic limit) normal
diffusion takes place even in the 'pen-case' model. When
collisions do not preserve momentum, $\kappa$ remains finite and
sound-wave effects are absent.

\end{abstract}
\maketitle
According to the Fourier's law of heat conductivity, when a small
temperature difference is applied across a system, in the steady
state the heat flux $j$ satisfies the equation
\begin{equation}
j=-\kappa \nabla T \label{fourier}
\end{equation}
where $T$ is a local temperature and $\kappa$ is the heat
conductivity of a given material. Since the heat conductivity
$\kappa$ is one of the most important transport coefficients a lot
of efforts were devoted to its calculation. Of particular interest
is the case of low-dimensional systems, where $\kappa$ might
diverge and thus Eq.(\ref{fourier}) would break down~\cite{LEPRI}.
However, despite intensive efforts the nature of this divergence
is not yet fully understood even in one-dimensional systems.
Renormalization group calculations~\cite{NARAYAN} show that for
one-dimensional fluid-like systems $\kappa$ should diverge with
the number of particles $N$ as $\kappa\sim N^{\alpha}$ with
$\alpha=1/3$. Although earlier works suggested other values of
$\alpha$, recent simulations for hard-core particle systems agree
with this prediction~\cite{GRASSBERGER}. In the other class of
systems, chains of nonlinear oscillators (Fermi-Pasta-Ulam
systems), simulations suggest~\cite{LEPRI,PROSEN} a larger value
of $\alpha$ (0.37-0.40) and that might be consistent with the
predictions of mode-coupling theory
$\alpha=2/5$~\cite{LEPRI1998,WANG2004}. However, recently it was
suggested that for such chains of oscillators $\alpha$ also should
be equal to 1/3  and numerically observed values of $\alpha$ were
attributed to numerical difficulties~\cite{MAI}.

It would be desirable to relate the divergence of heat
conductivity to other dynamic properties of a given system. For
example, it has been shown that chaoticity plays an important role
and ensures that $\kappa$ remains finite~\cite{ALONSO}. On the
other hand, conservation of momentum is known to imply the
divergence of $\kappa$~\cite{PROSEN,NARAYAN}. Some attempts were
also made to relate heat conductivity and diffusion. In
particular, Li and Wang suggested~\cite{LI2003} that the exponent
$\beta$ describing the mean square displacement of diffusing
particles
\begin{equation}
\langle x^2(t) \rangle \sim t^{\beta} \label{diff}
\end{equation}
should be related with $\alpha$ through the equation
\begin{equation}
\alpha=2-2/\beta. \label{liwang}
\end{equation}
Such an equation implies that the normal diffusion ($\beta=1$)
leads to the normal (non-divergent) heat conductivity
($\alpha=0$). Moreover, superdiffusion ($\beta>1$) and
subdiffusion ($\beta<1$) correspond to divergent ($\alpha>0$) and
vanishing ($\alpha<0$) heat conductivity,
respectively~\cite{comment1}. Although some numerical examples
~\cite{LIWANGWANG} seem to confirm the relation (\ref{liwang}),
its derivation is based only on qualitative arguments that
neglect, for example, interactions between particles and so the
suggestions that the relation (\ref{liwang}) is of more general
validity should be taken with care~\cite{METZLER}. In another
attempt, studying a class of noninteracting billiard heat
channels, Denisov \emph{et al.}~\cite{DENISOV} obtained a
different relation between the exponents $\alpha$ and $\beta$,
namely
\begin{equation}
\alpha=\beta -1. \label{denisov}
\end{equation}
The relation (\ref{denisov}) was verified numerically for the
energy diffusion in a one dimensional hard-core
model~\cite{DENISOV1}. However, it was argued~\cite{DENISOV1} that
the Levy walk scenario, that the energy diffusion obeys in this
model, might be due the absence of exponential instability and it
is not clear whether this result can be extended to more realistic
systems.

Establishing a firm relation between heat conductivity and
diffusion could be possibly very influential and  shed some light
also on other transport phenomena. In the present paper we examine
a model of hard disks in a narrow channel. When a fraction of
disks is immobile and thus collisions do not conserve momentum,
the heat conductivity $\kappa$ remains finite and normal diffusion
takes place. Simulations show that when there are no immobile
disks and momentum is conserved, heat conductivity diverges with
$\alpha$ close to 1/3. According to (\ref{liwang}) or
(\ref{denisov}) it should imply a superdiffusion. Although in this
case diffusion is enhanced by sound-wave effects, there are no
indications of superdiffusivity. Our work suggests that in
hard-disk systems in a narrow channel heat conductivity and
diffusion might be related but in a more intricate way than
relations (\ref{liwang}) or (\ref{denisov}) would suggest.

\begin{figure}
\vspace{2cm} \centerline{ \epsfxsize=15cm \epsfbox{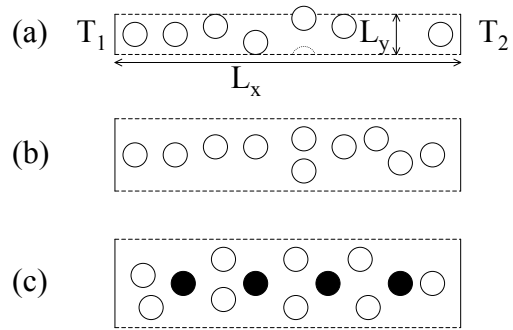} }
\vspace{-6cm} \caption{In our model disks move between thermal
walls kept at different temperatures $T_1$ and $T_2$. In vertical
direction periodic boundary conditions are used. In (a) particles
cannot pass each other and in (b) such a movement is possible. In
(c) a fraction of disks is immobile and placed in the middle of
the system (filled circles).} \label{fig1}
\end{figure}

In our model $N$ identical hard disks of radius $r$ and unit mass
are moving in a channel of size $L_x$ and $L_y$ (see
Fig.~\ref{fig1}). The channel is narrow ($L_x>>L_y$) and at both
of its ends there are thermal walls that are kept at temperatures
$T_1$ and $T_2$~\cite{TEHVER}. After the collision with the wall
kept at temperature $T$ a disk has its normal component sampled
from the distribution $p(v_x)=\frac{\Theta(\pm v_x)v_x}{T}{\rm
exp}(-\frac{v_x^2}{2T})$ with the sign in the argument of the
Heaviside function depending on the location of the wall. Its
parallel component is sampled from the Gaussian distribution
$p(v_y)=\frac{1}{\sqrt {2\pi T}}{\rm exp}(-\frac{v_y^2}{2T})$. In
the vertical $y$-direction periodic boundary conditions are used.
To calculate the heat conductivity $\kappa$ we use
Eq.(\ref{fourier}) with heat flux in the $x$-direction defined as
a time average of $j=\sum_i v_{xi}^3/2$ where $v_{xi}$ is the
$x$-component of the velocity of the $i$-th particle.

One can note that without heat reservoirs models of this kind are
chaotic~\cite{SINAI,SZASZ}. Such a feature makes this model more
realistic than, for example, a frequently examined one-dimensional
alternating-mass hard-disk model~\cite{GRASSBERGER,DHAR}. In the
simplest setup of our model, known as a 'pen case'
model~\cite{SINAI,SZASZ} (Fig.\ref{fig1}a) disks are so large that
they cannot exchange their position ($L_y/2>r>L_y/4$). Heat
conduction was already studied in such a case by Deutsch
 and Narayan~\cite{DEUTSCH}. Although their calculations indicate that
 $\alpha$ is close to 1/3, reported strong finite size effects and relatively small
 size of examined systems ($N\leq 2048$) suggest that one has to be cautious in the
 interpretation of these results. When $r$ and $L_y$ become small and the
 surface of particles is
considered as very rough the 'pen case' model becomes the
random-collision model. Calculations in such a case also suggest
that $\alpha$ is close to 1/3~\cite{DEUTSCH}.

An efficient way to simulate hard-disk systems is to use
event-driven molecular dynamics~\cite{RAPAPORT,BORIS,CORDERO}.
Performance of the algorithm considerably increases upon
implementing heap searching and sectorization and such methods
have already been applied to a number of
problems~\cite{LUDING,LIPOWSKI}. In our model it is sufficient to
use sectorization only in $x$ direction. With such a technique we
examined systems of up to $N=3\cdot 10^4$ hard disks. In the
'pen-case' version of the model there is no need to introduce
sectorization and simulations are only a little bit more demanding
than in a one-dimensional alternating-mass
model~\cite{GRASSBERGER}.

To allow an exchange of particles we simulated also systems with
disks of a smaller radius (Fig.\ref{fig1}b) and in such a case the
sectorization considerably speeds up simulations. To examine the
role of momentum conservation a fraction $c$ of disks are made
immobile. These disks are of the same radius and collisions with
them conserve energy but not momentum. They are placed along the
line $y=L_y/2$ (Fig.\ref{fig1}c) but similar (not presented)
results are obtained for the random distribution of immobile
disks. In our simulations the parameters were chosen as follows:
$L_x=N$, $L_y=1.0 - 1.5$, $r=0.01-0.3$, $T_1=1$, $T_2=2$, and
$c=0-0.1$. Initially centers of disks are usually uniformly
distributed along the line $y=L_y/2$ (for $c>0$
 they are between immobile disks).
 Their velocities are sampled from the Boltzmann distribution at
temperature interpolating linearly between $T_1$ at $x=0$ and
$T_2$ at $x=L$~\cite{comment2}. Such a system  evolves until a
stationary state is reached and then computations of some time
averages are made.

\begin{figure}
\centerline{ \epsfxsize=9cm \epsfbox{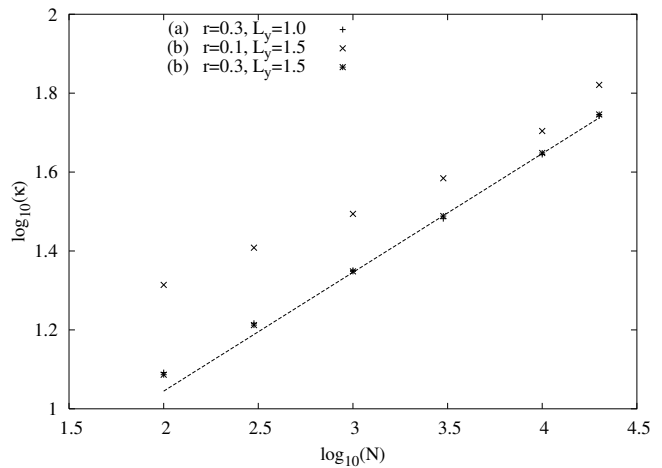} }
\caption{The size dependence of the heat conductivity $\kappa$ for
the momentum conserving cases (a) and (b). The solid line has a
slope corresponding to $\alpha=1/3$. For $r=0.3$ our results for
$L=1.0$ and $1.5$ are nearly identical.} \label{cond1}
\end{figure}

Numerical simulations in the momentum conserving case show that
$\kappa$ diverges with the number of particles $N$
(Fig.\ref{cond1}) and the exponent $\alpha$ is close to the
expected for one dimensional systems value $1/3$. In the limit of
ideal gas ($r\rightarrow 0$) heat flux is independent on $N$ (note
that $L_x=N$ and increasing the number of particles $N$ we also
increase the distance they had to travel) This explains the slower
convergence seen for $r=0.1$.

The divergence of $\kappa$ is an expected feature of momentum
conserving systems~\cite{PROSEN,NARAYAN}, and some arguments
suggest that in systems where momentum is not conserved $\kappa$
should be finite~\cite{NARAYAN}. A simple way to introduce
momentum nonconservation into our system is to place some immobile
disks. Although we do not present numerical data, our simulations
confirm that in such a case $\kappa$ remains finite in the
thermodynamic limit $N\rightarrow\infty$. We noticed that even a
small fraction of immobile disks ($c=0.01$) is sufficient to
remove the divergence of $\kappa$. It means that the behaviour of
$\kappa$ in a very sensitive way depends on the conservation of
momentum.

Simulations show that when there are no immobile disks, $\kappa$
diverges with the same exponent $\alpha(\approx 1/3)$ both when
particles cannot exchange their positions (case (a) in
Fig.\ref{fig1}) and when they can (case b). If Eq.(\ref{liwang})
or Eq.(\ref{denisov}) holds, we should observe in both cases the
superdiffusive behaviour (albeit with different exponents
$\beta$). To examine diffusive properties we measured the mean
square displacement $\langle x^2(t)\rangle$ over disks that at a
certain time enter the central part of the system and did not hit
a wall before the time $t$ (for the examined time scale such
processes were extremely rare)~\cite{comment3}. To measure
$\langle x^2(t)\rangle$ the system is not subjected to the
temperature difference ($T_1=T_2=T$).

\begin{figure}
\centerline{ \epsfxsize=9cm \epsfbox{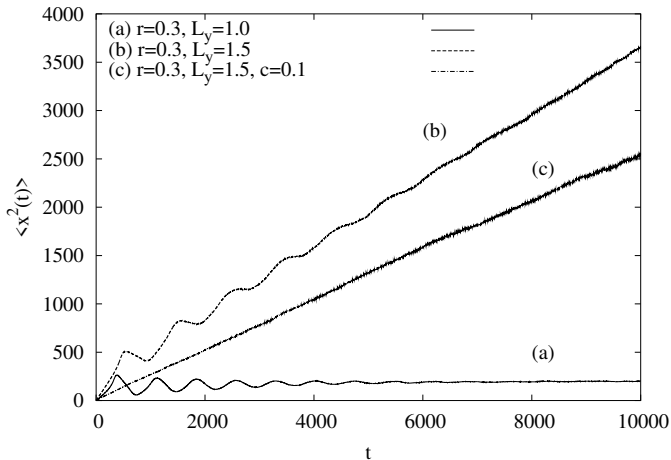} }
\caption{The time dependence of the m.s.q.~displacement $\langle
x^2(t) \rangle$ for $N=1000$ and $T_1=T_2=1$.} \label{dif1}
\end{figure}

Numerical results show (Fig.\ref{dif1}) that in the presence of
immobile disks $\langle x^2(t)\rangle$ increases linearly in time
($\beta=1$) and since $\kappa$ remains finite in this case
($\alpha=0$), both (\ref{liwang}) and (\ref{denisov}) are
satisfied. When momentum is conserved, two different behaviours
are observed. In the case (a) disks cannot exchange their
positions and displacement of particles is severely restricted. As
a result $\langle x^2(t)\rangle$ saturates and that is an
indication of strong subdiffusion. We will see, however, that this
is only a finite size effect and in the thermodynamic limit a
different behaviour emerges in this case. In the case (b) disks
can exchange their positions and asymptotically $\langle
x^2(t)\rangle$ increases linearly in time, as in the momentum
nonconserving case.

\begin{figure}
\centerline{ \epsfxsize=9cm \epsfbox{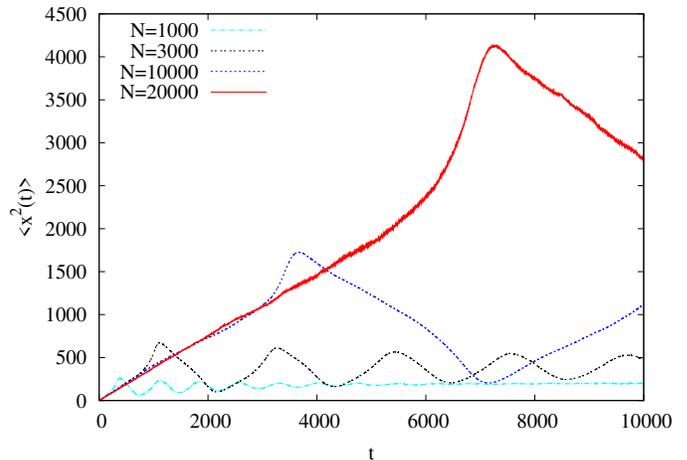} }
\caption{The time dependence of the m.s.q.~displacement $\langle
x^2(t) \rangle$ for momentum conserving case (a) with different
values of $N$ and $T_1=T_2=1.0$.} \label{difnarrow}
\end{figure}

Now, let us examine an interesting similarity in the behaviour of
$\langle x^2(t)\rangle$ in the case (a) and (b). Namely, initially
$\langle x^2(t)\rangle$ has some oscillatory behaviour in these
cases and there is no indication of such a behaviour in the
momentum nonconserving case. To examine the origin of these
oscillations we simulated systems of different number of disks and
at different temperatures. Simulations show that the time of the
first maximum of $\langle x^2(t)\rangle$ is approximately
proportional to $N$ (and thus to $L_x$) and inversely proportional
to $\sqrt{T}$ (note that $\sqrt{T}$ is proportional to the typical
velocity of disks). Such a behaviour indicates that
quasioscillations of $\langle x^2(t)\rangle$ are related with
sound-wave effects. Size dependence of these quasioscillations for
the case (a) is shown in Fig.\ref{difnarrow} and a similar
behaviour was found for the case (b). Let us also notice that in
case (b) the short-time growth of $\langle x^2(t)\rangle$  is
faster than the long-time growth, although in both cases the
growth is linear. Moreover, the saturation value of $\sqrt
{\langle x^2(t)\rangle}$ is much smaller than the system length
$L_x$.

Although at the large time scale the behaviour of $\langle
x^2(t)\rangle$ in cases (a) and (b) is much different
(Fig.\ref{dif1}), at the shorter time scale  it shows some
similarity.  Heat conductivity in cases (a) and (b) also behaves
similarly. It is thus tempting to suggest that sound-wave effects,
that provide a relatively fast transfer of energy but only on a
short time scale, are related both with divergence of $\kappa$ and
with quasioscillations of $\langle x^2(t)\rangle$. In presence of
immobile disks such effects disappear, apparently due to the
dissipation of momentum during collisions with immobile disks. As
a result $\kappa$ remains finite and $\langle x^2(t)\rangle$
increases monotonously in time. However, more detailed studies
would be needed to substantiate such a claim.

Let us also notice that the time scale set by sound-wave effects
diverges in the limit $N\rightarrow\infty$. In that case such
effects will dominate diffusive behaviour for arbitrarily long
time. As seen in Fig.\ref{difnarrow}, in such a limit $\langle
x^2(t)\rangle$ seems to develop longer and longer linear increase.
Thus we expect that in the limit $N\rightarrow\infty$, the
diffusion in both cases (a) and (b) (data not shown) is normal,
contrary to the predictions of (\ref{liwang}) or (\ref{denisov}).

That in the case (a) the mean square displacement $\langle
x^2(t)\rangle$ increases linearly in time is perhaps interesting
on its own. In this case particles cannot exchange their positions
and that resembles the molecular diffusion e.g., in some
zeolites~\cite{HAHN}. For such, so-called single-file systems, the
mean square displacement is known to increase as $\sqrt {t}$ and
such a slow increase was derived for some stochastic lattice gas
models~\cite{KUTNER}. As shown in Fig.\ref{difnarrow}, continuous
dynamics and/or momentum conservation considerably modify such a
behaviour. However, since the thermal motion is usually rather
fast, the sound-wave time scale is quite short and it might be
difficult to examine such effects experimentally.

In conclusion, our work shows that in momentum-conserving hard
disk systems in narrow channels heat conductivity $\kappa$
diverges with the exponent $\alpha\approx 1/3$ and sound-wave
effects enhance diffusion. As a result, in the thermodynamic
limit, normal diffusion appears even in the 'pen-case' version of
our model. When momentum is not conserved, $\kappa$ remains finite
and no enhancement of diffusion was observed. Heat conduction,
diffusion, and momentum conservation are terms of fundamental
importance in physics. Their intricate relations even in such
simple systems like the ones examined in the present paper should
warrant further study of these problems.

Acknowledgments: We gratefully acknowledge access to the computing
facilities at Pozna\'n Supercomputing and Networking Center.


\begin{thebibliography}{}
\bibitem{comment2} Such a choice of initial
velocities is relatively close to the the steady-state
distribution. However, the temperature profile in the steady state
is not exactly linear (see, e.g.,~\cite{GRASSBERGER,DHAR} for some
discussion of this aspect).
\bibitem{comment3} To measure $\langle x^2(t)\rangle$ some changes
in the program had to be made. In particular, with each disk that
enters the central part of the system we associate a separate
counter that measures the time that elapsed since a given disk
entered that central part.
\end{thebibliography}
\end {document}